\documentclass{PoS}

\title{Beta decay studies of r-process nuclei at the National Superconducting Cyclotron Laboratory}

\ShortTitle{Beta decay studies of r-process nuclei at the NSCL}

\author{\speaker{J.~Pereira}$^{a,b}$, A.~Aprahamian$^{c,d}$, O.~Arndt$^{e,f}$, A.~Becerril$^{a,b}$, T.~Elliot$^{a,b}$, A.~Estrade$^{a,b,g}$, D.~Galaviz$^{a,b}$, S.~Hennrich$^{e,f,a}$, P.~Hosmer$^{a,b,g}$,R.~Kessler$^{e,f,a}$, K.-L.~Kratz$^{f,h}$, G.~Lorusso$^{a,b,g}$, P.F.~Mantica$^{a,i}$, M.~Matos$^{a,b}$, F.~Montes$^{a,b,g}$, P.~Santi$^{a,b}$, B.~Pfeiffer$^{f,h}$, M.~Quinn$^{c,d}$, H.~Schatz$^{a,b,g}$, F.~Schertz$^{e,f,a}$, L.~Schnorrenberger$^{a,b,j}$, E.~Smith$^{b,k}$, B.E.~Tomlin$^{a,i}$, W.~Walters$^{l}$, A.~W\"{o}hr$^{c,d}$\\
\llap{$^{a}$}National Superconducting Cyclotron Laboratory, Michigan State University, E.~Lansing, MI, USA\\
\llap{$^{b}$}Joint Institute for Nuclear Astrophysics, Michigan State University, E.~Lansing, MI, USA\\
\llap{$^{c}$}Institute of Structure and Nuclear Astrophysics, Department of Physics, University of Notre Dame, South Bend, IN, USA\\
\llap{$^{d}$}Joint Institute for Nuclear Astrophysics, University of Notre Dame, South Bend, IN, USA\\
\llap{$^{e}$}Institut f\"{u}r Kernchemie, Universit\"{a}t Mainz, Mainz, Germany\\
\llap{$^{f}$}Virtuelles Institut f\"{u}r Struktur der Kerne and Nuklearer Astrophysik, Mainz, Germany\\
\llap{$^{g}$}Department of Physics and Astronomy, Michigan State University, E.~Lansing, MI, USA\\
\llap{$^{h}$}Max Planck Institut f\"{u}r Chemie, Otto-Hahn-Institut, Mainz, Germany\\
\llap{$^{i}$}Department of Chemistry, Michigan State University, E.~Lansing, MI, USA\\
\llap{$^{j}$}Institut f\"{u}r Kernphysik, TU Darmstadt, Darmstadt, Germany\\
\llap{$^{k}$}Department of Physics, Ohio State University, Columbus, OH, USA\\
\llap{$^{l}$}Department of Chemistry and Biochemistry, University of Maryland, College Park, MD, USA\\
}

\abstract{The impact of nuclear physics on astrophysical r-process models is discussed, emphasizing the importance of $\beta$-decay properties of neutron-rich nuclei. Several r-process motivated $\beta$-decay experiments performed at the National Superconducting Cyclotron Laboratory are presented. The experiments include the measurement of $\beta$-decay half-lives and neutron emission probabilities of nuclei in regions around $^{78}$Ni$_{50}$; $^{90}$Se$_{56}$; $^{106}$Zr$_{66}$ and $^{120}$Rh$_{74}$, as well as spectroscopic studies of $^{120}$Pd$_{74}$. A summary on the different experimental techniques employed, data analysis, results and impact on model calculations is presented.}

\FullConference{10th Symposium on Nuclei in the Cosmos\\
		 July 27 - August 1 2008\\
		 Mackinac Island, Michigan, USA}

\begin{document}

\section{Introduction}~\label{sec:Introduction}
Since its first appearance in the scientific literature~\cite{B2FH,Cam57}, more than fifty years ago, the rapid (r-) neutron-capture process remains one of the most exciting and challenging questions in nuclear astrophysics. In particular, the theoretical quest to explain the production of r-process isotopes, and the astrophysical scenario where this process occurs have not yet been satisfactorily solved (for a general review see, for instance, Ref.~\cite{Tru02,Cow06}). R-process isotopic and isobaric abundance distributions are typically deduced by subtracting the calculated s- and p- process contributions to the observed Solar System abundances. Furthermore, isotopic abundances originated in the early Galaxy can be directly observed in metal-poor Eu-enriched halo stars (MPEES) (i.e. $[Fe/H]$$\lesssim$$-2$, $[Ba/Eu]$$\lesssim$$-0.7$, $[Eu/Fe]$$\gtrsim$$+1$)
These combined observations reveal disparate behavior for light and heavy nuclei: MPEES abundance-patterns are nearly consistent from star to star and with the relative solar system r-process abundances for the heavier neutron-capture elements A$\gtrsim$130 (Ba and above), suggesting a rather robust main r-process operating over the history of the Galaxy. Such a consistent picture is not seen for light neutron-capture elements in the range 39$\leq$Z$\leq$50, as the Solar System Eu-normalized elemental abundances in MPPES show a scattered pattern~\cite{Joh02,Sne03,Aok05,Bar05}. Studies based on anti-correlation trends between elemental abundances and Eu richness at different metalicities provide evidence for a second different origin of isotopes below A$\simeq$130~\cite{Tra04,Mon07}. Measured abundances of $^{107}$Pd and $^{129}$I, for A$\textless$130, and $^{182}$Hf, for A$\textgreater$130$\--$trapped in meteorites in the Early Solar System formation~\cite{Rey60,Lee95}$\--$further reinforce the idea of different origins for isotopes lighter and heavier than A=130.

Different models are currently used to infer plausible sites for the nucleosynthesis of r-process-like neutron-rich elements. As an example, the high entropy wind model (HEW) can reproduce the abundance pattern over the entire mass regime using a unique core-collapse type II supernovae scenario during its different phases~\cite{Kra07}. Conversely, calculations based on the neutrino-driven wind model require three different sites: low-mass supernovae from $\sim$8-11~$M_{\odot}$ progenitors, normal supernovae ($\sim$12-25~$M_{\odot}$), and hypernovae ($\sim$25-50~$M_{\odot}$)~\cite{Qia08}. Besides the underlying astrophysical conditions, these models differ on the nuclear physics inputs used in the calculations. As an example, some authors claim that the calculated abundances in the region A$\simeq$$110$, prior to the A=130 peak$\--$wherein r-process models tend to underestimate the productions by an order of magnitude or more$\--$can be corrected by neutrino post-processing effects~\cite{Qia03}. On the other hand, calculations made on the basis of the classical r-process have shown that the underproduction of abundances can be corrected under the assumption of a reduction of the N=82 shell gap far from stability~\cite{Kra93,Pea96}. Clearly, an experimental confirmation of this quenching effect would have crucial consequences in the search for the r-process site. In this sense, self-consistent mean-field model calculations predict that such N=82 shell quenching might be associated with the emergence of a harmonic-oscillator doubly magic nucleus $^{110}$Zr, arising from the weakening of the energy potential surface due to neutron skins~\cite{Dob94,Pfe03}.

It is well known that $\beta$-decay properties play a crucial role in r-process model calculations~\cite{Kra93}: $\beta$-decay half-lives ($T_{1/2}$) of r-process waiting-point nuclei determine the pre freeze-out isobaric abundances and regulate the speed of the process towards heavier elements. $\beta$-delayed neutron emission probabilities ($P_{n}$) define the decay path towards stability during freeze-out, and provide a source of late-time neutrons. Besides astrophysical interest, these $\beta$-decay properties are related to the $\beta$-strength function $S_{\beta}(E)$, integrated over different energy ranges, and weighted by its strong energy dependence (relative to the $Q_{\beta}$ value) through the Fermi phase-space function $f(E)\sim (Q_{\beta}-E)^{5}$. Given these dependencies, $T_{1/2}$ and $P_{n}$ can be used as first probes of the structure of $\beta$-decay daughter nuclei in regions where more detailed structure studies are prohibitive owing to the low production rates and the lack of spectroscopic references for identification purposes. In particular, the $\beta$-decay properties of nuclei in the neighborhood of $^{110}$Zr may be exploited to investigate the spherical character of this potentially doubly-magic nucleus~\cite{Dob94}, or even a more exotic tetrahedral-symmetry type predicted by some authors~\cite{Sch04}. Complementary to these "gross" probes, $\beta$-delayed $\gamma$-spectroscopy has proven to be a very powerful tool to study the structure of nuclei in less exotic regions. Studies of the energy-level systematics as a function of neutron number serve to infer the nuclear structure effects that dominate the physics of r-process nuclei in more exotic regions.

The present paper summarizes the experimental activities carried out at the National Superconducting Cyclotron Laboratory (NSCL) aimed at studying $\beta$-decay properties of neutron-rich nuclei involved in the synthesis of heavy elements. The paper is divided in two sections: In the first part, the different experimental setups and measurement techniques are discussed. A second section follows, which summarizes the main results obtained in the different experimental campaigns, including the nuclear-structure insights inferred from the measurements.

\section{$\beta$-decay experiments at NSCL}
Exotic neutron-rich nuclei are produced and analyzed at NSCL, Michigan State University (MSU). The primary beam, accelerated in the Coupled Cyclotron Facility (CCF) at few hundred MeV per nucleon, impinges onto a production target (typically Be) with average intensities ranging from 1.5~pnA to 20~pnA. Fragmentation reaction products are forward-emitted and separated in-flight with the A1900 fragment separator~\cite{Mor03}, operated in its achromatic mode, using the $B\rho$$-$$\Delta E$$-$$B\rho$ technique~\cite{Sch87}. Two plastic scintillators located at the intermediate (dispersive) focal plane and
in the experimental area are used to measure the time-of-flight ($ToF$), related to the mass of the transmitted nuclei.

An Al wedge is mounted at the intermediate focal plane to keep the achromatism of the A1900; energy-loss experienced by nuclei passing through this degrader system provides a further filter to select a narrow group of elements. The selected cocktail-beam is transmitted into the experimental vault located downstream the A1900. Measured energy-losses ($\Delta E$) of these nuclei in a silicon PIN detector allow to further identify these nuclei according to their atomic number $Z$. The particle identification of the separated nuclei is verified by measuring $\gamma$-lines of well-known $\mu$s-isomers included in the cocktail-beam, in coincidence with $\Delta E$ and $ToF$ signals.


Finally, the separated and identified nuclei reach the experimental area where they are analyzed.

\subsection{$\beta$-decay half-lives $\--$ NSCL Beta Counting System}
The Beta Counting System (BCS)~\cite{Pri03} consists of a stack of silicon detectors, the center piece being a 40$\times$40-pixel doubly-sided 1~mm-thick Si strip detector (DSSD), where implanted nuclei are detected, together with their position-correlated subsequent $\beta$-decays. Detection efficiencies of $\beta$-decays are typically better than 30$\%$. Downstream of the DSSD, a stack of 16-strip single sided Si detectors (SSSD) and a 1~cm-thick Ge crystal are used to veto light particles and nuclei that punch-through the implantation detector, and for calorimetry of $\beta$ particles. The output signals of the DSSD and SSSD detectors are coupled to dual-gain preamplifiers designed to deal with the wide dynamic energy range associated with implanted heavy nuclei (GeV) and emitted $\beta$-decays (keV).
The whole system is enclosed in a cylindrical, thin walled Al vacuum chamber, coupled to the beam line.

Implanted nuclei are unambiguously identified according to $\Delta E$, $ToF$ and total kinetic energy ($TKE$), which is determined as the sum of energy-loss signals over all silicon detectors. The time and pixel where an implantation event is detected are recorded. Subsequent decay events observed in the same or neighboring pixels within a specified time window are assumed to be correlated. The distribution of time differences between implantation and decay events provides the corresponding $\beta$-decay curve for a given nucleus. Multi-parameter least-square fits are calculated for these curves using the Batemann equations for mother, daughter and granddaughter decays, as well as background. Half-lives of daughter and granddaughter are fixed (when known), whereas background, initial number of parent nuclei and their decay half-lives are left as free parameters. Alternatively, a maximum-likelyhood (MLH) analysis can be used in cases where Gaussian statistics are not realized. The MLH method has been extensively used in previous analysis where the low production rates of some very exotic nuclei precluded a fit analysis of their $\beta$-decay curves based on the Batemann equation~\cite{Hos05,Mon06,Ber90,Sch95}.

\subsection{$\beta$-delayed neutron-emission probabilities $\--$ The Neutron Emission Ratio Observer}
Neutrons emitted after the $\beta$ decay of a nucleus are measured at NSCL with the Neutron Emission Ratio Observer (NERO)~\cite{Lor06}. The detector consists of forty-four BF$_{3}$ and sixteen $^{3}$He proportional gas counters, embedded in a polyethylene matrix with a central hole that accommodates the BCS chamber along its symmetry axis. $\beta$-delayed neutrons are moderated in the polyethylene block and subsequently detected by the proportional gas counters.
Signals from these counters are registered in a multi-hit TDC during a 200~$\mu$s window following $\beta$ decay; this time interval was chosen on the basis of an average moderation time of about 150~$\mu$s measured for neutrons emitted from a $^{252}$Cf source. The $P_{n}$ value of a nucleus is deduced from the ratio of the number of detected $\beta$-delayed neutrons$\--$corrected for background contamination and limited efficiencies around 40$\%$$\--$to the number of $\beta$ decays.

\subsection{$\beta$-delayed $\gamma$-spectroscopy $\--$ The Segmented Germanium Array}
Another detector widely used in $\beta$-decay experiments at NSCL is the Segmented Germanium Array (SeGA)~\cite{Mue01}. Sixteen detectors from SeGA can be mounted in compact geometry around the BCS. The $\gamma$-peak detection efficiency is 7$\%$ and the energy resolutions 3.5~keV at 1~MeV. $\beta$ decays from the nuclei implanted in the DSSD can populate the excited states of the daughter nuclei. De-exciting $\gamma$ rays are detected with SeGA whenever a trigger signal from the $\beta$ decay is registered. The combination of SeGA with the BCS provides a mean to dissentangle the level ordering of excited states in the daughter nucleus. The $\beta$ branching ratios derived from $\gamma$ ray intensities can facilitate the assignment of quantum numbers ($J^{\pi}$) to the excited states.

\section{Experimental results and discussion}
\label{results}

\subsection{Half-life of the doubly-magic r-process waiting point $^{78}$Ni}
\begin{figure*}
\begin{center}
\includegraphics[width=12cm]{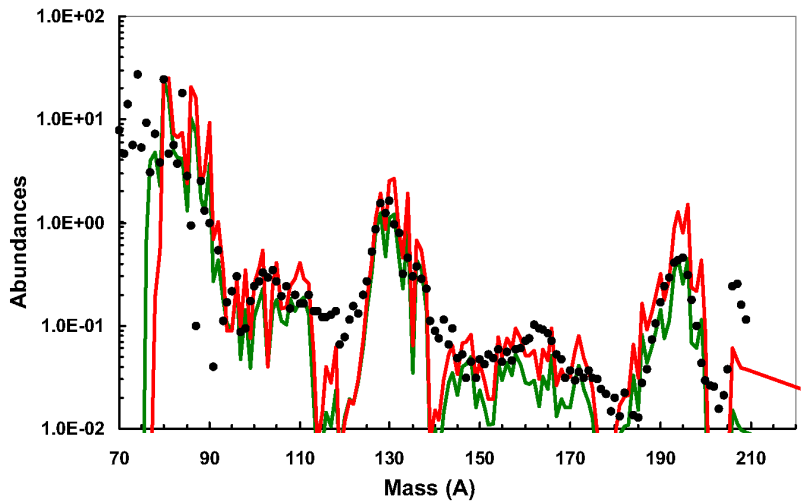}
\caption{(Color Online) Solar System r-process isotopic abundances (dots) compared with two classical r-process calculations. The first one was obtained using QRPA calculated half-lives for every nucleus (green line); the second one was obtained after replacing the calculated half-life of just $^{78}$Ni by the new measured value (red line). Figure based on Ref.~\cite{Hos08}}
\label{fig:78Nirprocess}
\end{center}
\end{figure*}

One of the main achievements at NSCL, regarding r-process motivated experiments, was the measurement of the half-life of the doubly-magic $^{78}$Ni nucleus, along with half-lives and $\beta$-delayed neutron branchings of neighboring isotopes~\cite{Hos05}. The new measured half-life, $T_{1/2}=110^{+100}_{-60}$~ms, is significantly shorter than that obtained from QRPA-type calculations~\cite{Mol90,Mol97}. $^{78}$Ni represents an important waiting point in r-process models where the neutron-captures begin at rather light masses, such as those assuming non standard neutron star masses~\cite{Ter01} or those triggering the supernovae explosion with the collapse of an ONeMg core~\cite{Wan03}. Classical r-process calculations aimed at reproducing light neutron-capture elemental abundances (39$\leq$$Z$$\leq$50) may also be directly affected by this waiting point. In these scenarios, the flow of matter through the N=50 shell is constrained by the half-life of $^{78}$Ni, along with those of the other two already-known waiting points $^{79}$Cu and $^{80}$Zn, and has a strong impact on the formation and shape of the abundance pattern beyond the A=80 peak. Indeed, classical r-process calculations have shown that the replacement of the half-life of $^{78}$Ni, obtained from P.~M\"{o}ller's QRPA model~\cite{Mol97}, by the new (more than four times shorter) measured value has strong impact on the synthesis of heavy r-process nuclei, with particular emphasis for the cosmochronometers actinides Th and U (see Fig.~\ref{fig:78Nirprocess}). Much better agreement of the calculated half-life with the experimental value is found in the updated QRPA-model version discussed in Ref.~\cite{Mol03}.
of neutrons during phases where the very r-process is not yet fully developed.

\subsection{$\beta$-decay half-lives and $P_{n}$ values of nuclei approaching the N=82 r-process path}
The half-lives of $^{111,115}$Tc, $^{114-118}$Ru, $^{116-121}$Rh and $^{119-124}$Pd, and $P_{n}$ values of $^{116-120}$Rh, $^{120-122}$Pd and $^{124}$Ag, near the r-process path, were also measured at NSCL~\cite{Mon06}. The region $A$=112$-$124 is particularly interesting as it lies in the well-known abundance trough$\--$albeit none of the measured nuclei are part of the r-process path itself. The measured half-lives were compared with QRPA calculations~\cite{Mol90} using ground-state deformations, and $Q_{\beta}$ and $S_{n}$ predicted by both, the ETFSI-Q model~\cite{Pea96}, which includes quenching effects of the N=82 shell in a phenomenological way, and the FRDM model~\cite{Mol95}. Although neither of these inputs provided a complete agreement with the data, better results were obtained when ETFSI-Q was used instead of FRDM. Such result may indicate a weakening of the N=82 shell closure for Pd isotopes with N$\geq$74, although other parameters entering into the calculation of the $\beta$-strength function should be better known. As for $P_{n}$ values, QRPA calculations using mass extrapolations and ETFSI-Q ground-state deformations show good agreement for most of the cases, with $^{120}$Rh being an exception. For this nucleus, a discrepancy of a factor 4 is found between $P_{n}$ values calculated based on ETFSI-Q masses, and the experimental limit of $P_{n}\simeq 5\%$. This discrepancy could be due to uncertainties in the quadrupole deformation and $Q_{\beta}$ values used in the calculations. When the new $P_{n}$ value is included in a classical r-process calculations, the $^{120}$Sn/$^{119}$Sn abundance ratio increases by 40$\%$.

\subsection{Collectivity of Pd isotopes approaching the N=82 shell: $\beta$-delayed $\gamma$-spectroscopy of $^{120}$Pd}
The structure of nuclei in the vicinity of the classical r-process path are crucial for testing the theoretical models aimed at calculating the relevant properties of r-process nuclei experimentally inaccessible. These studies are particularly important in the abundance trough region of the r-process mass abundances. Following this motivation, an experiment aimed at investigating the collectivity of Pd even-even neutron-rich isotopes via $E(2^{+}_{1})$ and $E(4^{+}_{1})/E(2^{+}_{1})$ measurements was performed at NSCL~\cite{Wal04}. As the low-lying yrast levels are expected to rise when approaching a closed shell, deviations from such a trend may reveal an enhanced collective character for these nuclei.

\begin{figure*}
\begin{center}
\includegraphics[width=6cm]{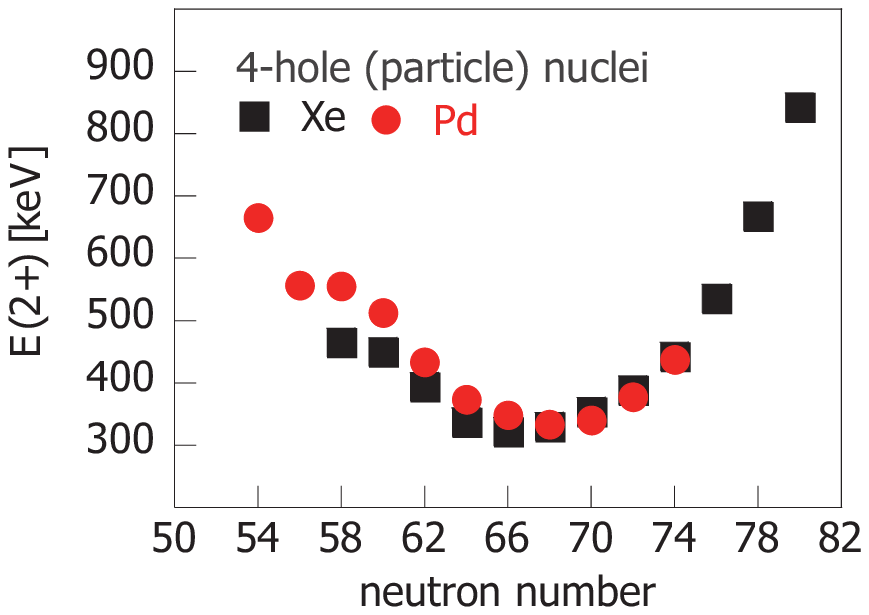}
\includegraphics[width=6cm]{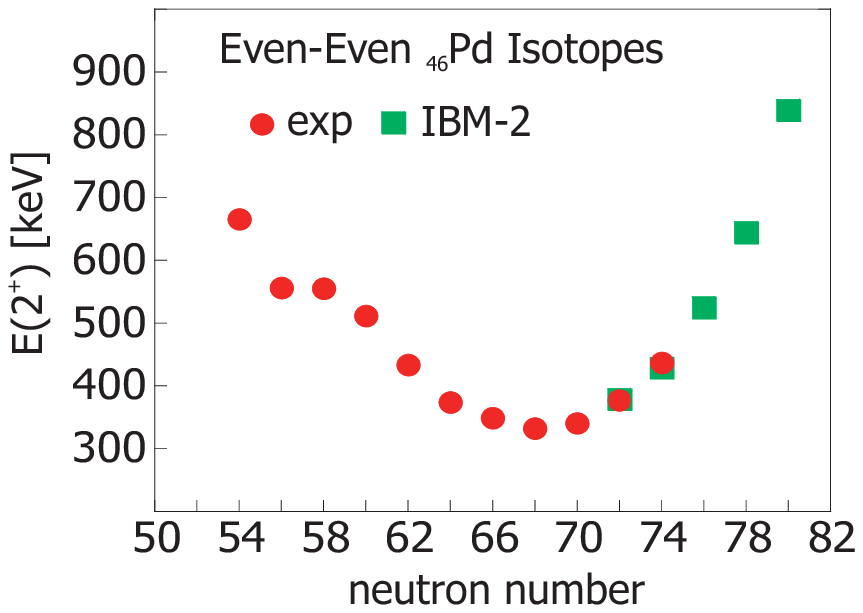}\\
\caption{(Color Online) Measured E($2^{+}_{1}$) systematics of even-even Pd isotopes (circles) compared with (left) 4-proton particle isotones $_{54}$Xe (black squares), and (right) IBM-2 calculations (green squares). (Figures taken from Refs.~\cite{Wal04,Tom06}).}
\label{Pd}
\end{center}
\end{figure*}

$\gamma$ rays with energies of 438~keV and 618~keV were observed in coincidence with $\beta$ particles following decay of $^{120}$Rh$_{75}$. These transitions were assigned as the $2^{+}_{1} \rightarrow 0^{+}_{1}$ and $4^{+}_{1} \rightarrow 2^{+}_{1}$, respectively, in the daughter $^{120}$Pd$_{74}$. The identified $E(2^{+}_{1})$ level of this nucleus, along with its even-even neighbor $^{118}$Pd$_{72}$, lie in near proximity of those corresponding to $^{108}$Pd$_{62}$ and $^{110}$Pd$_{64}$, revealing isotopic symmetry centered in $^{114}$Pd$_{68}$ (see Fig.~\ref{Pd}, left). Likewise, a regular trend was observed when comparing those nuclei with their corresponding 4 proton-particle isotones $^{126}$Xe$_{72}$ and $^{128}$Xe$_{74}$.

As shown in Fig.~\ref{Pd} (right), earlier IBM-2 $E(2^{+}_{1})$ systematics$\--$calculated on the basis of boson number established from N=82 and Z=50 closed shells$\--$reproduce well the whole range of Pd isotopes, including the measured value in $^{120}$Pd. This result supports the notion that the collectivity of Pd isotopes is systematically reduced beyond the $N\simeq$$66$ mid-shell~\cite{Wal04}, which may have important consequences for the calculations of the mass and $\beta$-decay properties of the (potential) waiting point $^{128}$Pd$_{82}$.

\subsection{$\beta$-decay properties of r-process nuclei around N$\approx$66 mid-shell}
Two other recent experiments at NSCL include the study of N$\approx$66 mid-shell nuclei below A=110, and neutron-rich As and Se isotopes at the N$\approx$56 sudden onset of deformation. In these experiments, new half-lives were measured for the first time for $^{88,89}$As, $^{90}$Se, $^{105}$Y, $^{106,107}$Zr and $^{111}$Mo, and new $P_{n}$ values were determined for $^{103,104}$Y, $^{104}$Zr and $^{109,110}$Mo. In addition, upper limits could be established for the $P_{n}$ values for $^{105}$Y, $^{103,105-107}$Zr and $^{108,111}$Mo.

\begin{figure*}[h]
\begin{center}
\includegraphics[width=6cm]{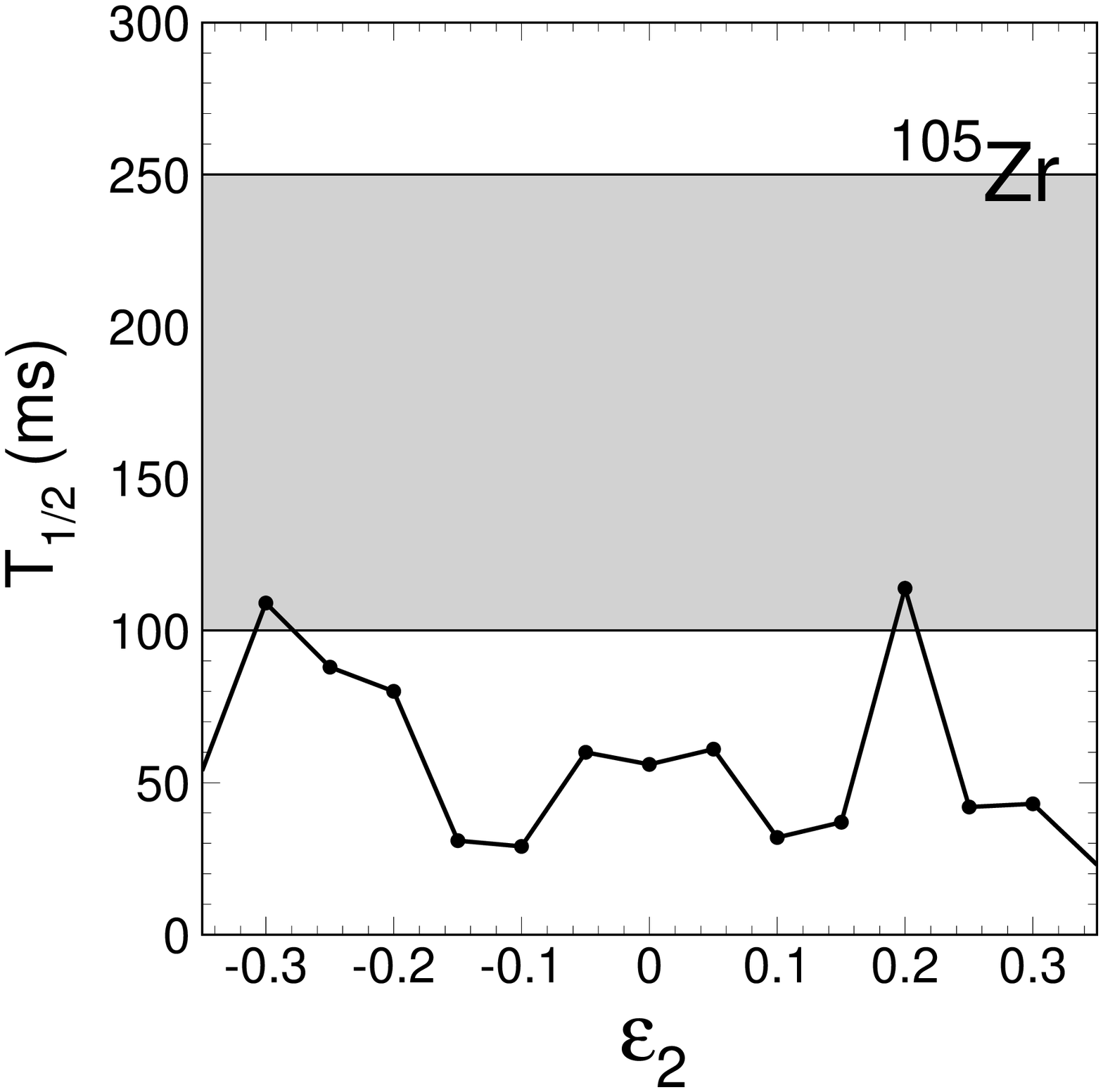} 
\includegraphics[width=6cm]{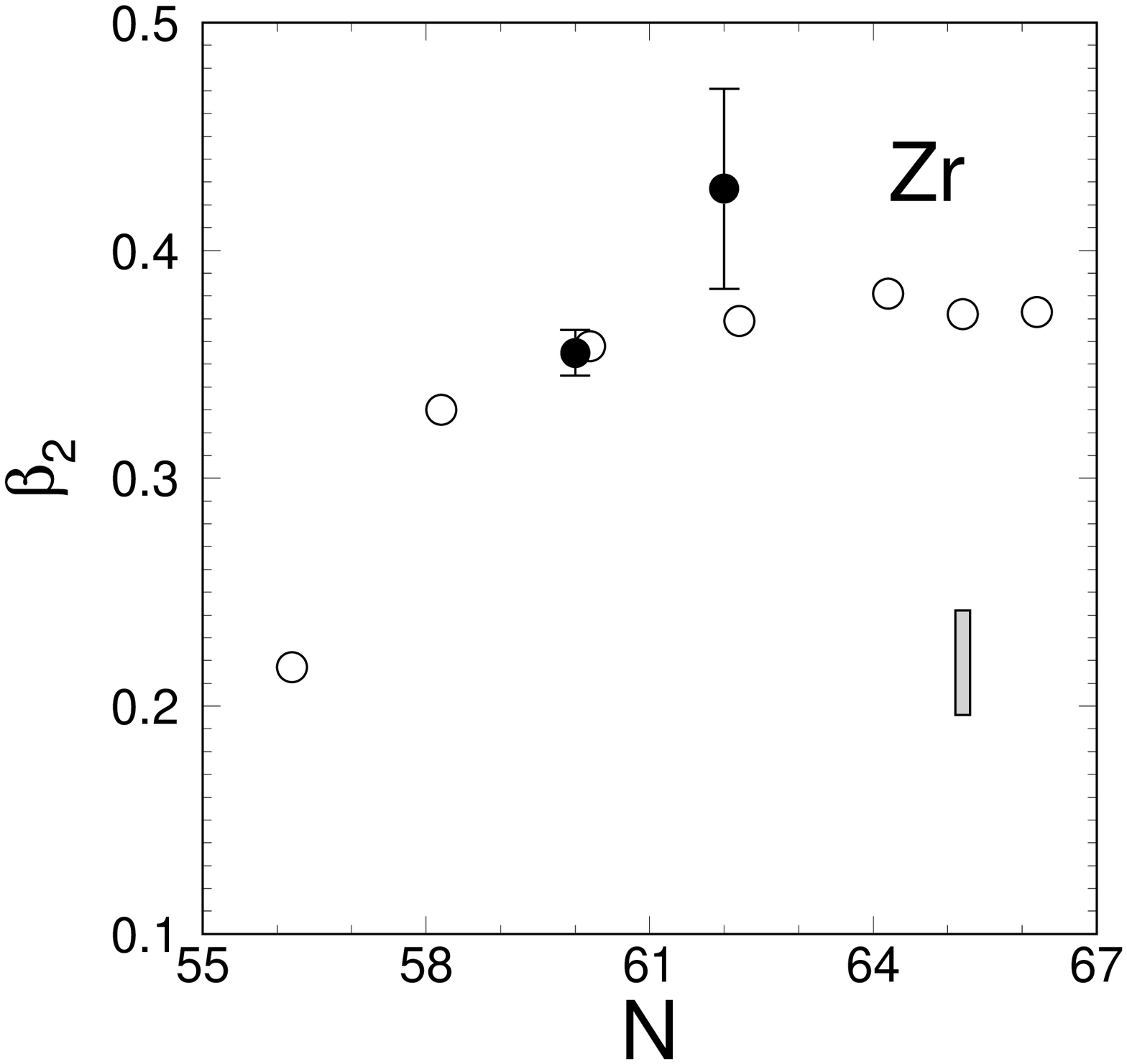} 
\caption{Left: QRPA-calculated $T_{1/2}$ for $^{105}$Zr as a function of $\epsilon_{2}$ (solid line), compared with the measured value (shaded area). Right: Quadrupole deformation parameter $\beta_{2}$ of Zr isotopes obtained from: S.~Raman \emph{et al.}~\cite{Ram01} (solid dots), FRDM calculations (empty dots), QRPA-optimized half-life (shaded rectangle).}
\label{fig:Zrdef}
\end{center}
\end{figure*}

From the measurements of mid-shell isotopes in the vicinity of Zr it was possible to investigate the systematics of quadrupole deformation parameters $\beta_{2}$ with neutron number. The measured $T_{1/2}$ and $P_{n}$ were calculated with the QRPA model using mass extrapolations and $\epsilon_{2}$ varied over a wide range of deformations. As an example, Fig~\ref{fig:Zrdef} (left) shows the calculated $T_{1/2}$ of $^{105}$Zr, as a function of $\epsilon_{2}$, compared with the new experimental data.

The range of prolate deformations $\beta_{2}$ (derived from $\epsilon_{2}$), that optimized the QRPA results to best reproduce the measured $T_{1/2}$ of $^{105}$Zr$_{65}$, is compared in Fig.~\ref{fig:Zrdef} (right) with the values predicted by the FRDM model. Interestingly, the deformation of Zr isotopes, expected to reach saturation for isotopes at the mid-shell, gets its maximum value around $^{104}$Zr$_{64}$. Beyond this isotope, the evolution of $\beta_{2}$ with neutron number deviates from the FRDM predictions, showing a more pronounced decrease. This result may indicate a shape transition region, from highly-prolate to spherical shapes, due to a possible double shell closure at Z=40, N=70, as predicted by J.~Dobaczewski \emph{et al.}~\cite{Dob94}. According to these authors, the expected doubly-magic character of $^{110}$Zr$_{70}$ arises from the weakening of the energy potential surface due to neutron skins near the drip line, which would be directly related with a quenching of the N=82 shell~\cite{Dob94,Pfe03}. Alternatively, other authors attribute the low deformations in this region to the presence of a tetrahedral-symmetric potential minimum at $^{110}$Zr$_{70}$~\cite{Sch04}. Further detailed $\beta$-delayed $\gamma$-spectroscopic studies, necessary to clarify such possibility, will have to wait for the development of new high-intensity fragmentation-beam facilities based on FRIB concept.

\section{Conclusions}
R-process motivated $\beta$-decay campaigns have been undertaken at the NSCL in recent years. The high-intense beams from the NSCL Coupled Cyclotron Facility, along with the high resolution achieved by the A1900 separator, were crucial for producing the very exotic species near and at the r-process path. Furthermore, the versatility of the BCS, NERO and SeGA detection systems enabled the measurement of $\beta$-decay properties important to r-process modeling. $\beta$-decay half-lives were deduced using two independent methods: the least-square fitting of the decay-curves and the Maximum Likelihood technique. Delayed neutrons and $\gamma$ rays detected in coincidence with $\beta$ particles provided a clean method to extract $P_{n}$ values and to analyze the low-energy spectra of the daughter nuclei.

The half-life of the doubly-magic $^{78}$Ni nucleus was measured for the first time. Classical r-process calculations have shown that the inclusion of the short new half-life ($T_{1/2}=110^{+100}_{-60}$~ms) of the waiting-point $^{78}$Ni has a strong impact on the synthesis of heavy r-process nuclei, with particular emphasis for the cosmocronometers actinides Th and U.

New $T_{1/2}$ of $^{115}$Tc, $^{116-118}$Ru, $^{121}$Rh and $^{121-124}$Pd; $P_{n}$ values of $^{118,119}$Rh, $^{124}$Pd and $^{124}$Ag; and upper $P_{n}$ limits of $^{116,117,120}$Ru and $^{120-122}$Pd were also measured. QRPA-calculated half-lives of Pd isotopes show better agreement with the new measured data when ground-state deformations, $Q_{\beta}$ and $S_{n}$ are taken from ETFSI-Q. Furthermore, classical r-process calculations have shown that the abundance ratio $^{120}$Sn/$^{119}$Sn is strongly affected by the measured $P_{n}$ of $^{120}$Rh. $\beta$-$\gamma$ spectroscopy of $^{120}$Rh provided first information on low-energy yrast states in $^{120}$Pd. The systematic variation of $E(2^{+}_{1})$ and $E(4^{+}_{1})$ in the even-even Pd isotopes up to $^{120}$Pd$_{74}$ show good agreement with IBM-2 calculations which assume good shell closure at Z=50 and N=82.

New half-lives were also measured for $^{88,89}$Ar, $^{90}$Se, $^{105}$Y, $^{106,107}$Zr and $^{111}$Mo, along with $P_{n}$ values of $^{103,104}$Y, $^{104}$Zr and $^{109,110}$Mo, and upper $P_{n}$ limits of $^{105}$Y, $^{103,105-107}$Zr and $^{108,111}$Mo. QRPA calculations using $\beta_{2}$ parameters optimized to reproduce the new measured half-life of $^{105}$Zr$_{65}$ show a reduction of the quadrupole deformation for these nuclei. Such trend may indicate the presence of a spherically symmetric $^{110}$Zr$_{70}$ predicted by some authors.

Extension of $\beta$-decay measurements to full r-process pathway requires new high-intensity fragmentation-beam facilities like FRIB. In the mean time, fully microscopic $\beta$-decay models are necessary to make reliable extrapolations of $\beta$-decay properties to r-process nuclei.

\end{document}